\documentclass[conference]{IEEEtran}
\IEEEoverridecommandlockouts
\usepackage{cite}
\usepackage{amsmath,amssymb,amsfonts}
\usepackage{algorithmic}
\usepackage{graphicx}
\usepackage{subcaption}
\usepackage{textcomp}
\usepackage{xcolor}
\usepackage{bm}
\usepackage{amsthm}

\def\BibTeX{{\rm B\kern-.05em{\sc i\kern-.025em b}\kern-.08em
    T\kern-.1667em\lower.7ex\hbox{E}\kern-.125emX}}

\renewcommand\baselinestretch{0.94}

\IEEEaftertitletext{\vspace{-0.5cm}}

\begin{document}

\title{Neuromorphic Algorithm-hardware Codesign for Temporal Pattern Learning\\
}

\author{
\small
\IEEEauthorblockN{ Haowen Fang$^{1}$, Brady Taylor$^{2}$, Ziru Li$^{2}$, Zaidao Mei$^{1}$, Hai (Helen) Li$^{2}$, Qinru Qiu$^{1}$}
\IEEEauthorblockA{ $^{1}$Department of Electrical Engineering and Computer Science, Syracuse University, Syracuse, NY\\
$^{2}$Department of Electrical and Computer Engineering, Duke University, Durham, NC\\
Email: { $^1$\{hfang02, zmei05, qiqiu\}@syr.edu, $^{2}$\{brady.g.taylor, ziru.li, hai.li\}@duke.edu} }
}

\maketitle

\begin{abstract}
Neuromorphic computing and spiking neural networks (SNN) mimic the behavior of biological systems and have drawn interest for their potential to perform cognitive tasks with high energy efficiency. However, some factors such as temporal dynamics and spike timings prove critical for information processing but are often ignored by existing works, limiting the performance and applications of neuromorphic computing. On one hand, due to the lack of effective SNN training algorithms, it is difficult to utilize the temporal neural dynamics. Many existing algorithms still treat neuron activation statistically. On the other hand, utilizing temporal neural dynamics also poses challenges to hardware design. Synapses exhibit temporal dynamics, serving as memory units that hold historical information, but are often simplified as a connection with weight. Most current models integrate synaptic activations in some storage medium to represent membrane potential and institute a hard reset of membrane potential after the neuron emits a spike. This is done for its simplicity in hardware, requiring only a ``clear" signal to wipe the storage medium, but destroys temporal information stored in the neuron.

In this work, we derive an efficient training algorithm for Leaky Integrate and Fire neurons, which is capable of training a SNN to learn complex spatial temporal patterns. We achieved competitive accuracy on two complex datasets. We also demonstrate the advantage of our model by a novel temporal pattern association task. Codesigned with this algorithm, we have developed a CMOS circuit implementation for a memristor-based network of neuron and synapses which retains critical neural dynamics with reduced complexity. This circuit implementation of the neuron model is simulated to demonstrate its ability to react to temporal spiking patterns with an adaptive threshold.

\end{abstract}


\section{Introduction} \label{sec:intro}


The brain-inspired Spiking Neural Network (SNN) has drawn interest as it demonstrates high energy efficiency in performing machine learning tasks. In contrast to artificial neural networks, which can be generally formulated as matrix operations, SNN is a neural network with dynamics, and time is an essential computing element in its computation. Another unique feature of SNNs is that the neuron's output is a discrete spike event, and hence information in a SNN is delivered and processed as spike trains. How the information is encoded into spike trains is still an open question. Existing approaches can be roughly divided into rate coding models and temporal coding models. Rate coding models assume there is a time window, where spike number within the window resembles a continuous value in deep neural network (DNN). Inputs in each window are spike trains of particular rates that are processed window by window. Hence, a purely rate-based system suffers from significant latency and inefficiency when processing time-varying inputs \cite{gerstner2014neuronal,eliasmith2004neural}. Rate-based models only consider spike statistics inside each window, ignoring dependencies in spike trains and neuron states, which are proven to be critical in information processing \cite{eliasmith2004neural}.

In biological neural systems, an aforementioned window is hardly observed in which neurons fire at particular rates. Instead, SNNs are inherently dynamical systems dealing with time-varying stimulus. Inputs from physical environments change over time, and neurons' output spike patterns are time-varying correspondingly. Information is naturally embedded in the temporal structure of spike trains \cite{eliasmith2004neural}, referred to as temporal coding. Temporal coding models allow fast and accurate information processing and are therefore beneficial for low latency and real time applications, such as classifying sensor readings or generating control signals to respond to external stimulus \cite{eliasmith2004neural}. To exploit the potential of SNNs to process temporal information, a novel training algorithm is required. Recently, there have been works employing Backpropagation Through Time (BPTT) to train SNNs to learn temporal patterns \cite{zenke2018superspike, neftci2019surrogate}. However, these works use relatively complex neuron models and dynamics, posing challenges for deploying trained models on hardware. Therefore, a model with reduced complexity but still retains critical aspects of neural dynamics and a training algorithm that can train such a model to learn temporal information is highly desirable for hardware design. 

In the conventional neural network or neuromorphic computing hardware implementation, synaptic weights are stored in SRAM cells. The computations often involve accessing to SRAM, accumulating activations. The frequent data accessing and movement cause power and latency overhead. In addition, a SRAM cell can require six or more transistors for a single bit, consuming much power and area. Therefore, Resistive Random-access Memory (RRAM) has drawn much attention, as it enables integrated data storage and computation. A memristor-based memory can store one or more bits of a synapse weight in a single device. Furthermore, RRAM reads are extraordinarily quick and consume little power, and computations with these synaptic weights can be executed directly in the memory without the need to retrieve data~\cite{yang2013memristive}. These computations are carried out by applying voltages to the word-lines of the memory, generating proportional currents through each cell that are accumulated at each bit-line. While an DNN implemented in an RRAM-based system would require costly digital-to-analog and analog-to-digital converters to interface with the array, SNNs are uniquely compatible for these systems. Binary spike inputs generate currents that are integrated on capacitors at each bit-line, and an analog comparator spikes at a threshold, discharging the capacitor~\cite{liu2015spiking}. However, many existing works assumes rate-based model and ignores the neural dynamics such as \cite{wang2015energy}. Indeed, there are hardware designs that stress the realistic neural model such as BrainScaleS\cite{schmitt2017neuromorphic}. Such designs target at neuroscience research rather than machine learning tasks, they can replicate realistic neural dynamics however at the cost of additional design complexity. On one hand, oversimplified design limits the potential of SNN, on the other hand, the complexity introduced by implementing biologically realistic neural dynamics may not be necessary for machine learning tasks. This motivates us to find a sweet spot between hardware complexity and computational efficiency. That is, we aim at developing a hardware architecture that exhibits critical components of neural dynamic without introducing significant overhead.

In this work, our contributions are summarized as follows:
\begin{itemize}
    \item We derive a training algorithm for hardware friendly Leaky Integrate and Fire (LIF) spiking neuron model.
    \item The model allows us to process spatial temporal pattern without explicit recurrent network structure, thus reduces the complexity of crossbar based implementation. Model simplification and transformation are also adopted to facilitate hardware implementation.
    \item Though employing a simplified neuron model, our algorithm can still train SNNs to learn complex spatial temporal patterns. This enables novel applications of SNNs, rather than static image classification. 
    \item We achieve competitive accuracy on common SNN datasets : 98.40 \%, 85.69 \% on N-MNIST and Spiking Heidelberg Digits (SHD) datasets.
    \item We codesign a neuron circuit for RRAM-based architectures that replicates the dynamics of our hardware friendly neurosynaptic model.
\end{itemize}

\section{Model} \label{sec:model}

We start from a standard Leaky Integrate and Fire (LIF) model, which is defined by Ordinary Differential Equations (ODEs) and event-triggered update \cite{brette2007simulation}:

\vspace{-2mm}
\begin{subequations}
\label{eq:neuron_model}
\begin{gather}
      \tau \frac{dv(t)}{dt} = -v(t) + \sum_{i}^M w_i x_i(t)       \label{eq:ode_1} \\
      v(t) \leftarrow v_{rest} \text{, if  } v(t) = V_{th} \label{eq:ode_2} 
\end{gather}
\end{subequations}

where $v(t)$ is neuron's membrane potential, $\tau$ is a time constant, $x_i(t)$ is input of the $i_{th}$ synapse, $w_i$ is associated weight, $M$ is the synapse number, $V_{th}$ is threshold, $v_{rest}$ is rest potential, and for simplicity, we set $v_{rest} = 0$. When $v(t)$ exceeds $V_{th}$, $v(t)$ is set to $v_{rest}$, this is called "hard reset". It is obvious that the spiking neuron is a stateful system as $v(t)$ integrates input over time, however the temporal dependency is implicit in this model. Since our purpose is to utilize the temporal dynamics of SNNs, next we employ the Spike Response Model (SRM) framework\cite{gerstner2014neuronal} to derive an explicit temporal dependency. To simplify discussion let $M=1$, \eqref{eq:ode_1} describes a Linear time-invariant (LTI) system. An LTI system can also be characterized by impulse response. Solving \eqref{eq:ode_1} with initial condition $v(0) = 0$ and a pulse as input, we can obtain kernel $k(t) = e^{\frac{-t}{\tau}}$ \cite{gerstner2014neuronal}. For a neuron with $M$ input synapses, we have

\vspace{-2mm}
\begin{equation} \label{eq:psp}
    PSP(t) = \sum_i^M w_i \int_0^\infty e^{\frac{-t}{\tau}}x_i(t-s)ds
\end{equation}

$PSP(t)$ is referred as post-synaptic potential, it is the theoretical neuron potential without reset. SRM implements reset by a negative charge induced by output spike to neuron itself, which can also be characterized by a reset kernel $h(t)$. In principle, $h(t)$ can have various forms, and we choose $h(t) = e^{\frac{-t}{\tau_r}}$, where $\tau_r$ is a time constant. Therefore, \eqref{eq:ode_1} - \eqref{eq:ode_2} can be converted to following form:

\vspace{-0.2cm}
\begin{equation} \label{eq:srm}
\resizebox{0.90\linewidth}{!}{$
    \displaystyle
    v(t) = - \vartheta \int_{0}^{\infty} h(t) O(t-s)ds + \sum_i^M w_i  \int_{0}^{\infty} k(s)x_i(t-s) ds$}
\end{equation}

\vspace{-0.1cm}

where $O(t) = \sum \delta(t-t^f), t^f \in \{t^f: v(t^f) \geq V_{th} \}$ is a sequence of time-shifted Dirac Delta functions, representing neuron's output spike train, $\vartheta$ determines the strength of the reset charge. SRM has following advantages: 1) equation \eqref{eq:srm} establishes an analytical relation of temporal dependency, at any time $t$, $v(t)$ is determined by entire input and output history; 2) neuron is less likely to fire after reset, however, ODE model cannot quantitatively reflect such likelihood due to the hard reset, while $h(t)$ serves as a bridge between previous history and current output. 3) the hard reset simply set membrane potential to 0, impairing historical information. 

\begin{figure}[t]
  \centering
  \includegraphics[width=0.9\linewidth]{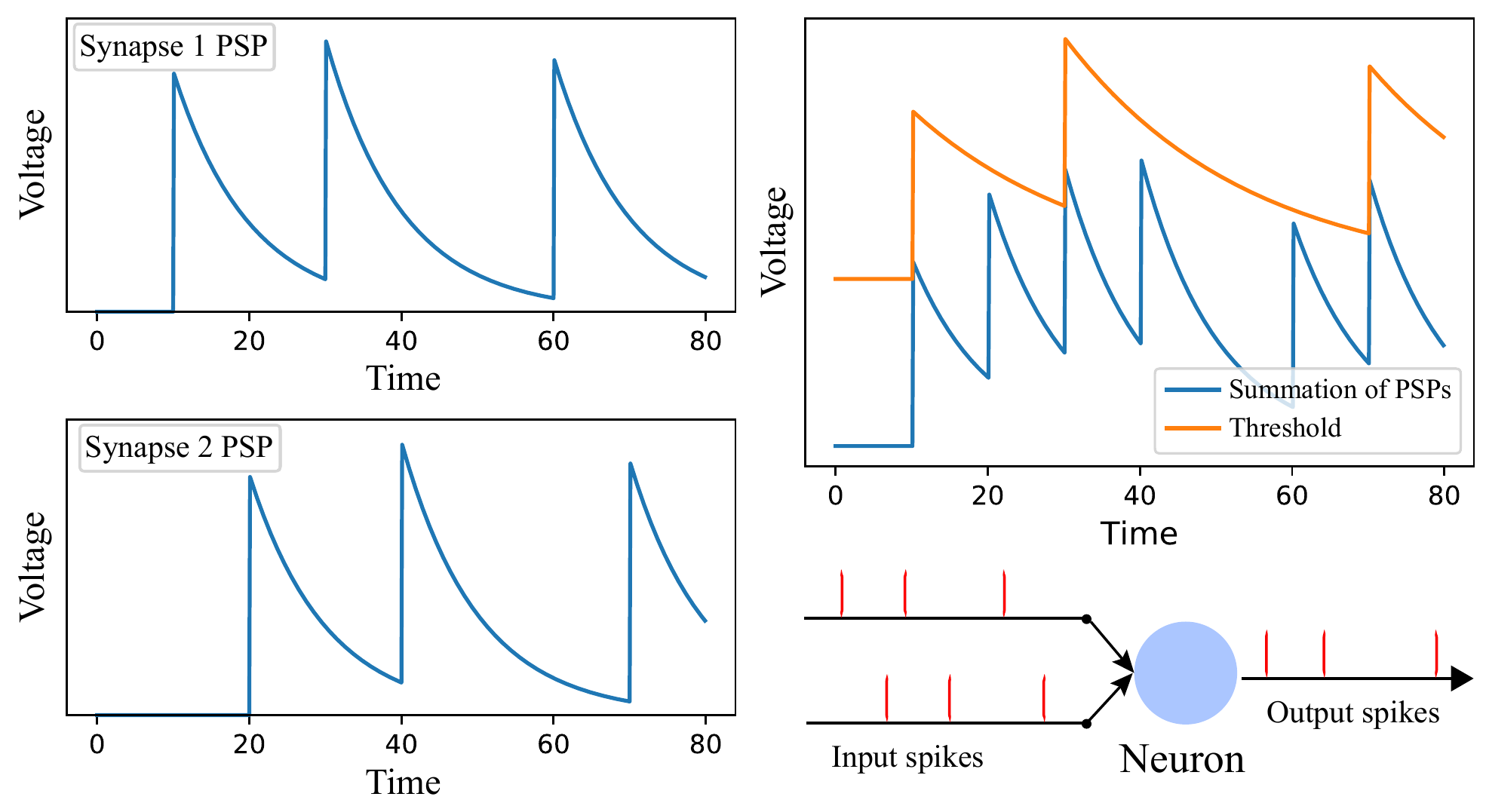}
  \caption{Synapse and adaptive threshold dynamic.}
  \label{fig:dynamic}
\end{figure}

One potential problem with equation \eqref{eq:srm} is that voltage subtraction is more difficult to realize than addition. However, resetting neuron by a negative charge is equivalent to adopting an adaptive threshold, such that voltage subtraction is avoided. We rewrite \eqref{eq:srm} as:

\vspace{-3mm}
\begin{subequations}
\label{eq:adaptive}
\begin{gather}
      PSP(t) = \sum_i^M w_i  \int_{0}^{\infty} k(s)x_i(t-s) ds    \label{eq:adaptive_1} \\
      r(t) = \vartheta \int_{0}^{\infty} h(t) O(t-s)ds \label{eq:adaptive_2} \\
      O(t) = \sum \delta(t-t^f), t^f \in \{t^f: v(t^f) \geq V_{th} + r(t) \} \label{eq:adaptive_3}
\end{gather}
\end{subequations}
where $r(t)$ is the amplitude of the reset charge. Instead of comparing $v(t)$ with a fixed threshold, in this form, $PSP(t)$ is compared with a time-varying threshold $V_{th} + r(t)$ which adaptively changes with respect to output spike activity. This leads to adaptive threshold circuit design to be detailed in Section \ref{sec:hardware}. Fig. \ref{fig:dynamic} illustrates the dynamic of synapse and threshold. The left two figures indicate the synapse PSP change with respect to input spikes. Blue curve in top right figure indicates summation of PSPs, and the orange curve is the threshold. After issuing a spike, the threshold increases immediately, and exponentially decays over time. The importance of such reset mechanism and drawback of ODE model with hard reset will be shown by experiment in Section \ref{sec:classification}.

Equation \eqref{eq:srm} also has a physical interpretation, which explains the neuron's capability to remember information and provides insights to circuit design. Essentially, $h(t)$ and $k(t)$ are two first order low-pass filters. A neuron with $M$ input synapses is expressed as $M + 1$ filters. Therefore, neuron's memory is distributed to filters, and the filter states are never cleared. Unlike in the ODE model \eqref{eq:neuron_model}, memory only exists in neuron, after a hard reset, historical information is discarded. A first order low-pass filter can be implemented by a RC circuit, with an input signal applied across a resistor and capacitor in series and an output signal measured across the capacitor. The values selected for these discrete components largely depends on the duration of a single time-step $\Delta t$ in the physically-realized algorithm. The time constant of a filter $\tau$ can be represented by the equation \(\tau = \frac{RC}{\Delta t}\). While recurrent structure causes additional complexity in memristor crossbar based non-digital design, the filter-based model allows us to process temporal sequence with a feedforward network, hence reduces difficulty in implementation.

\section{Training Algorithm} \label{sec:training}

Our next step is to derive training algorithm based on BPTT, it is necessary to get a step-based updating rule. By using Z-transform \cite{brette2007simulation}, we can obtain the digital counterpart of filter $k(t)$ and $h(t)$ in discrete time domain:

\vspace{-3mm}
\begin{subequations}
\label{eq:discrete_filter}
\begin{gather}
      k[t] = e^{\frac{-1}{\tau}} k[t-1] + x[t]       \label{eq:discrete_k} \\
      h[t] = e^{\frac{-1}{\tau_r}} h[t-1] + O[t-1] \label{eq:discrete_h}
\end{gather}
\end{subequations}

where $t \in \mathbb{Z}_{\ge0}$. Therefore, \eqref{eq:srm} can be transformed into a system of difference equations:

\vspace{-3mm}
\begin{equation} \label{eq:model_1}
    \boldsymbol{v}_{l}[t] = \boldsymbol{g}_{l}[t] - \vartheta \boldsymbol{h}_l[t]
\end{equation}
\vspace{-4mm}
\begin{equation} \label{eq:model_2}
    \boldsymbol{g}_{l}[t] = \boldsymbol{W}_l {\boldsymbol{k}_l[t]}
\end{equation}
\vspace{-4mm}
\begin{equation} \label{eq:model_3}
    \boldsymbol{h}_l[t] = e^{\frac{-1}{\tau_r}} \boldsymbol{h}_l[t-1] + \boldsymbol{O}_{l}[t-1] 
\end{equation}
 \vspace{-4mm}
\begin{equation} \label{eq:model_4}
    \boldsymbol{k}_l[t] = e^{\frac{-1}{\tau}} \boldsymbol{k}_l[t-1] + \boldsymbol{O}_{l-1}[t] 
\end{equation}
\vspace{-4mm}
\begin{equation} \label{eq:model_5}
    \boldsymbol{O}_{l}[t] = U(\boldsymbol{v}_{l}[t] - V_{th}) 
\end{equation}
\vspace{-4mm}
\begin{equation} \label{eq:model_6}
    U(x) = 0, x < 0 \text{ otherwise 1} 
\end{equation}

\begin{figure}[t]
  \centering
  \includegraphics[width=\linewidth]{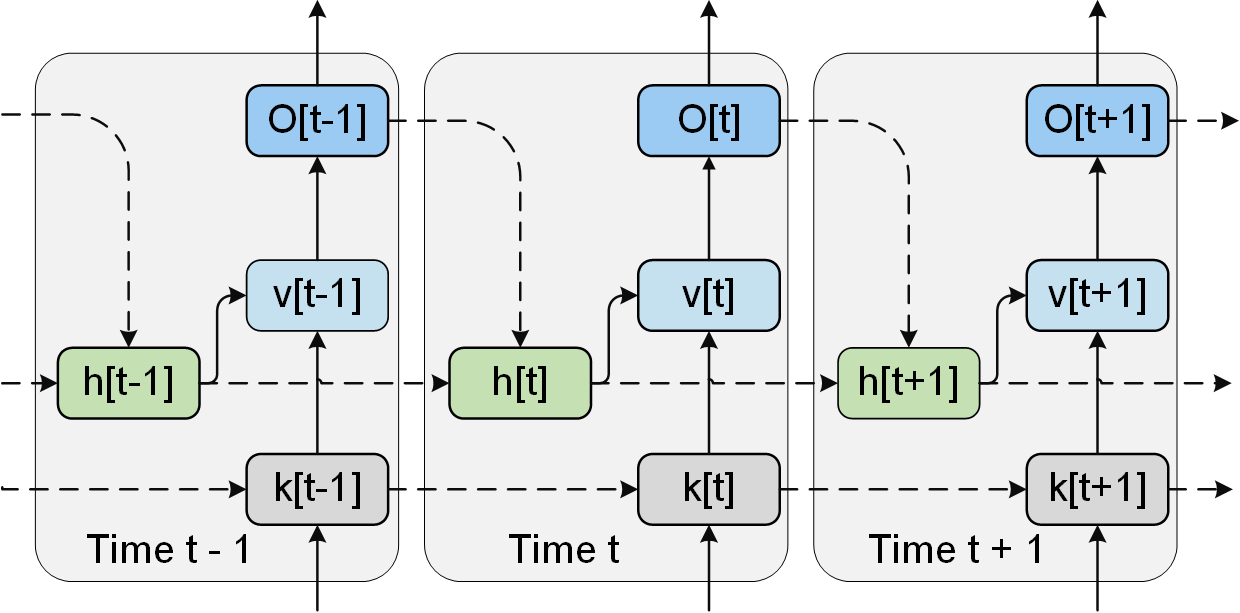}
  \caption{Unfolded network.}
  \label{fig:unfold}
\end{figure}

where $l$ denotes layer index, $\boldsymbol{W}_l \in \mathbb{R}^{ N_l {\times} N_{l {-} 1} }$ is weight matrix, $\boldsymbol{g}_{l}[t]$ is weighted input (PSP), $\boldsymbol{O}_l[t]$ is neuron output at time $t$, $U(x)$ is Heaviside step function. Like \eqref{eq:adaptive_1} to \eqref{eq:adaptive_3}, equation \eqref{eq:model_1} \eqref{eq:model_5} and \eqref{eq:model_6} can also be interpreted as adaptive threshold:

\vspace{-2mm}
\begin{equation}
    O[t] = 1 \quad  \text{if} \quad  g[t] > \vartheta h[t] + V_{th} \quad \text{else} \quad 0
\end{equation}
\vspace{-3mm}

Equation \eqref{eq:model_1} to \eqref{eq:model_6} exhibit recursive form, after unfolding BPTT can be applied. The unfolded network is shown in Fig. \ref{fig:unfold}. Let $E$ be the loss and $\boldsymbol{\delta}_{l}[t] = \frac{{\partial E}}{{\partial \boldsymbol{O}_l[t]}}$ be the error signal. $\boldsymbol{\delta}_{l}[t]$ can be propagated recursively:

\vspace{-3mm}
\begin{equation}
\resizebox{0.89\linewidth}{!}{$
    \displaystyle
    \boldsymbol{\delta}_{l}[t] = {\boldsymbol{W}^\intercal_{l {+} 1}} (\boldsymbol{\epsilon}_{l {+} 1}[t] \boldsymbol{\delta}_{l {+} 1}[t]) - \vartheta \boldsymbol{\delta}_l[t+1] \boldsymbol{\epsilon}_l[t+1] $
    }
\end{equation}


where $\boldsymbol{\epsilon}_{l}[t] = \frac{{\partial U(\boldsymbol{v}_{l}[t]-\vartheta)}}{{\partial \boldsymbol{v}_{l}[t]}}$. Equation \eqref{eq:model_1} to \eqref{eq:model_5} are differentialble, however derivative of $U(x)$ is a Dirac Delta function, which is a roadblock of backpropagation. We solve this issue by pseudo-gradient \cite{neftci2019surrogate}, such that $U'(x)$ is approximated by the directive of a complementary error function:

\vspace{-2mm}
\begin{equation}
  \frac{{\partial U(x)}}{{\partial x}} \approx \frac{ \partial( \frac{1}{2} \text{erfc}(\frac{x}{\sqrt{2}\sigma}))}{\partial x}   =  - \frac{\exp{(-\frac{x^2}{2\sigma^2}})}{\sqrt{2\pi}\sigma}
\end{equation}

where $\sigma$ determines the sharpness of derivative.

We consider two learning tasks. First is the conventional classification task, we use cross-entropy loss and spike rate is mapped to probability by Softmax function. In second task, we consider a non-trivial scenario. As discussed in section \ref{sec:intro}, SNN can generate particular temporal patterns in response to specific input patterns, which is common in control and memory system \cite{eliasmith2004neural,gerstner2014neuronal}. Therefore, in this task, we aim at training the network to generate output $O[t]$ which temporally resembles target spike train $S_{target}[t]$. In other words, the goal is to train network to generate spikes at specific times. The loss is defined as the distance between $O[t]$ and $S_{target}[t]$. Essentially, spike train is a sequence of time shifted Dirac Delta functions, it is difficult to directly measure the similarity. We employ kernel method \cite{park2013kernel} to map discrete spike train to continuous trace, such that the distance between two arbitrary spike trains $S_i[t]$ and $S_j[t]$ is calculated as:

\vspace{-2mm}
\begin{equation}
    D(S_i, S_j) = \frac{1}{2T} \sum_{t=1}^T (f[t]*S_i[t] - f[t]*S_j[t])^2
\end{equation}
\vspace{-2mm}

where $T$ is spike train length, $*$ denotes convolution and kernel function $f[t] = e^{\frac{-t}{\tau_m}} - e^{\frac{-t}{\tau_s}}$. The loss is the summation of distances between each output and target pairs:

\begin{equation}
E =  \sum_{i=1}^{N_L} D(O^L_i, S^i_{target})
\end{equation}
\vspace{-2mm}

where $i$ denotes output and target spike train index ,$N_L$ is number of output spike trains.

\section{Hardware Design} \label{sec:hardware}

\begin{figure}[t]
  \centering
  \includegraphics[width=\linewidth]{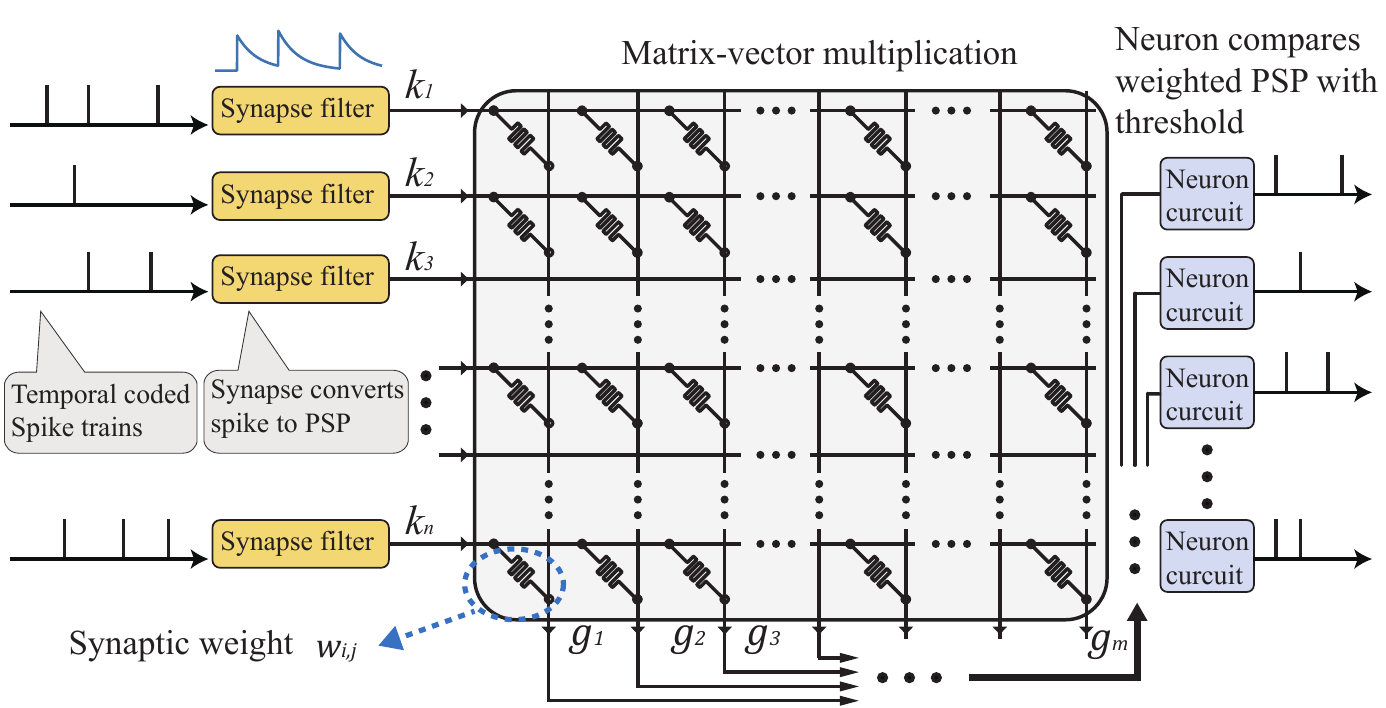}
  \caption{System structure.}
  \label{fig:structure}
\end{figure}

Equation \eqref{eq:srm} reveals the filter nature of spiking neurons. This provides insights to physical realization:

\begin{itemize}
    \item Biological plausible neural dynamic can be achieved by low-pass filters, reducing circuit design complexity. 

    \item Adaptive threshold, an important neural dynamic, can be implemented by a comparator and a low-pass filter.
\end{itemize}

Fig. \ref{fig:structure} shows the overall structure of our hardware design. An array of synapse filters receives time-varying spike trains as input, implementing filter $k(t)$. The output of synapse filter array is fed into crossbar to perform matrix-vector multiplication, corresponding to \eqref{eq:model_2}. The resulting PSPs are compared with threshold in neuron circuit. Output spikes of neuron circuit are sent to synapse filter array of next layer.

Implementation of proposed neuron model and synaptic dynamics is relatively simple. An incoming input spike is shaped by the RC filter into an exponentially decaying waveform. These voltage waveforms are applied to their respective word-lines to induce a current in each bit-line that can be interpreted as the dot-product of the bit-line synapse conductances and word-line voltages. These dot-products are the PSPs. At the edge of each bit-line is a single resistor that converts the output current of the bit-line to a voltage. While this resistor may have an effect on the resulting current, this can be avoided with the inclusion of a current amplifier between the output of the bit-line and the resistor as in~\cite{liu2016memristor}. We ignore this effect for the sake of this work as it should only affect the magnitude of the resulting current and not the shape. 

The generated PSP voltages ($k(t)$) are applied to the positive input of an operational amplifier, which acts as a comparator. The threshold voltage at the negative input is the the feedback of the comparator through a separate, identical RC filter ($h(t)$) plus a bias ($V_{th}$). The bias is added to $h(t)$ through the use of a separate operational amplifier that gives the comparator's negative input an offset. When $k(t)$ rises above the threshold, the comparator goes high, which in turn raises $h(t)$ and the threshold voltage, causing the amplifier to go low again and creating a spiking behavior. The output of the comparator is buffered to the output of the neuron by two inverters to ensure the magnitude of the output spikes is \(V_{DD}\). These spikes can then be buffered to the input of the next layer.

In the first RC filter, a voltage spike over a very short period of time is charging the capacitor to a voltage that will slowly decay. In order to ensure that the voltage reached by the capacitor produces an output within the operating range of the amplifier in the neuron circuit, a level-shifter can be used between layers to raise the input-spike amplitude above \(V_{DD}\). In order for the amplifier to drive the second RC filter, the amplifier must have a strong second-stage capable of driving the load. The circuit diagram for our neurosynaptic circuit with a single neuron and single synapse can be seen in Fig.~\ref{fig:circuit}.
  
\section{Experiments and Evaluations}

\begin{table}[t]
\caption{Parameters}
\centering
\label{tab:parameters}
\setlength\tabcolsep{4 pt}
\begin{tabular}{|c|c||c|c||c|c|}
\hline
Parameter  & Value & Parameter & Value & Parameter & Value \\ \hline
Optimizer & AdamW & Batch size &  64    & $\tau$       & 4     \\ \hline
\begin{tabular}[c]{@{}c@{}}Learning rate\\ (classification)\end{tabular} &  0.0001 & $\tau_r$ &  4 & $\tau_m$ &  4 \\ \hline
\begin{tabular}[c]{@{}c@{}}Learning rate\\ (pattern association)\end{tabular}     & 0.001  & $\sigma$      & $\frac{1}{\sqrt{2\pi}}$   &  $\tau_s$     & 1     \\ \hline
\end{tabular}
\end{table}

To demonstrate the effectiveness of our model and training algorithm, we applied them on classification and temporal pattern association tasks. The second is non-trivial as it requires network to recognize specific spatial temporal input patterns and also produce spike trains that follow particular patterns. Our model and training algorithm are implemented in PyTorch. Detailed parameters are given in Table \ref{tab:parameters}. Results of circuit level simulation is also provided and compared with the desired behavior.

\subsection{Spiking Dataset Classification} \label{sec:classification}

We tested on two spiking datasets and compared our neuron model with standard LIF model to demonstrate the proposed model's advantage in temporal pattern classification.

\begin{figure}[t]
  \centering
  \includegraphics[width=\linewidth]{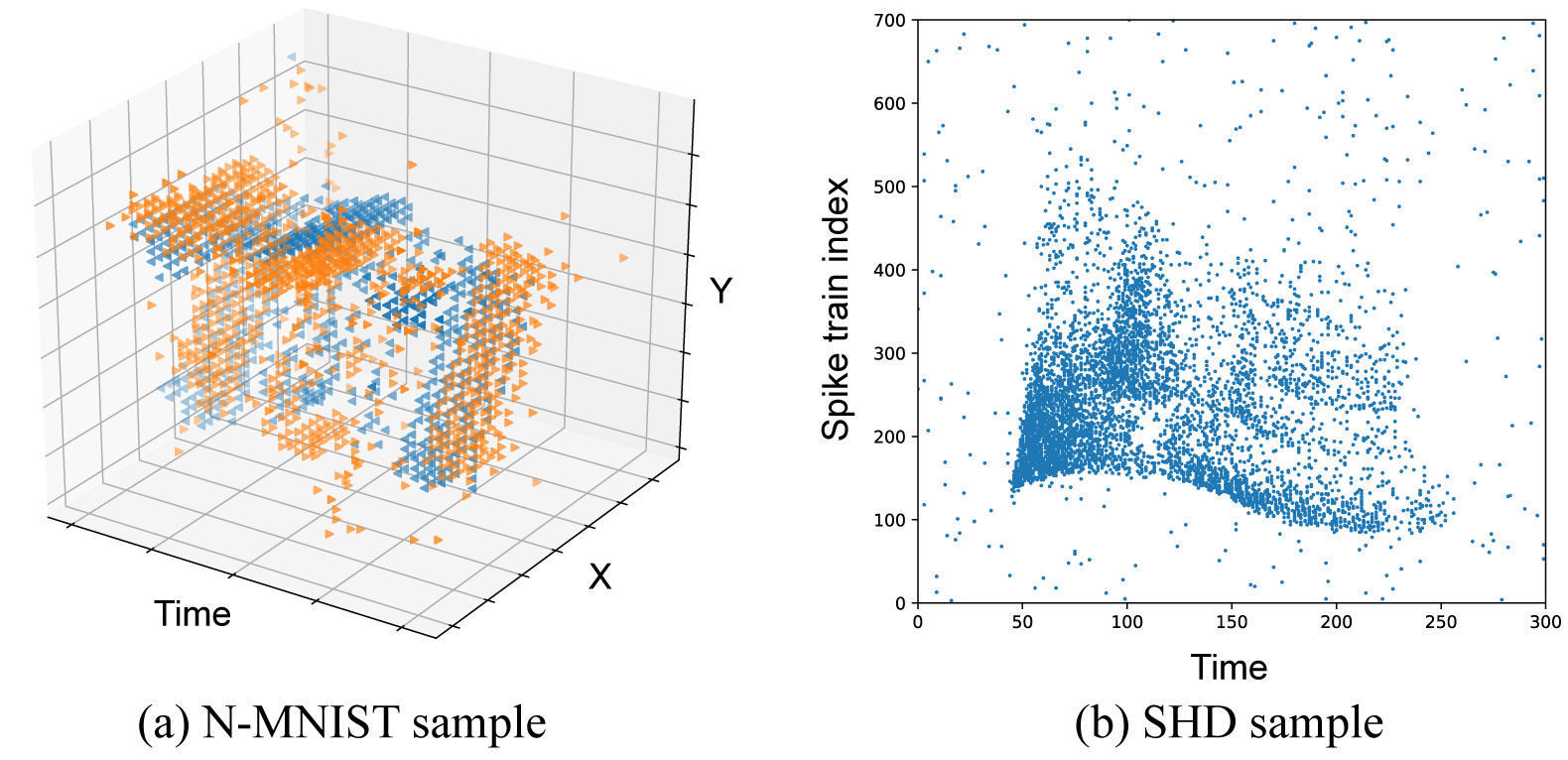}
  \caption{N-MNIST and SHD dataset samples.}
  \label{fig:samples}
\end{figure}

\begin{table}[t]
\centering
\caption{Classification Results}
\label{tab:classification}
\begin{tabular}{|c|c||c|c|}
\hline
\multicolumn{2}{|c||}{N-MNIST}                               & \multicolumn{2}{c|}{SHD}                      \\ \hline
Model       & Accuracy  & Model & Accuracy \\ \hline
This work                            & 98.40  &  This work         &  85.69   \\ \hline
This work (HR)                       & 95.31  &  This work (HR)    &   26.36  \\ \hline
Spiking MLP \cite{lee2016training}   & 98.66  &  Spiking MLP \cite{cramer2019heidelberg}     & 47.5     \\ \hline
Phased LSTM \cite{neil2016phased}    & 97.28  &  R-SNN \cite{cramer2019heidelberg}     &  83.2    \\ \hline
Spiking CNN \cite{neil2016effective} & 95.72  &  LSTM\cite{cramer2019heidelberg}     & 89.0     \\ \hline
Graph CNN \cite{bi2019graph}         & 98.5   &  R-SNN \cite{zenke2020remarkable}     &  82.0    \\ \hline
Spiking CNN \cite{vaila2019deep}     & 98.32  &  SRNN\cite{yin2020effective}   &   84.4   \\ \hline
\end{tabular}
\end{table}

\textbf{\emph{N-MNIST classification.}} Neuromorphic MNIST (N-MNIST) is the dynamic version of MNIST handwritten digits, captured by recording MNIST images displayed on LCD screen using a DVS camera mounted on a moving platform. Once the brightness change at position $(x,y)$ exceeds a certain threshold, a spike event is triggered. The data exhibits complex temporal patterns. A sample of N-MNIST demonstrating the spatial temporal spike distribution is shown in Fig. \ref{fig:samples} (a). The blue and orange dot represent spikes of two channels. Our network to classify N-MNIST is a Multi-Layer Perceptron (MLP), structure is $(34 {\times} 34 {\times} 2) {-} 500{-} 500 {-} 10$. Comparisons with existing works are shown in Table \ref{tab:classification}. We achieved $98.40 \%$ accuracy without more expensive training techniques such as error normalization in \cite{lee2016training}. We outperformed convolutional SNNs such as \cite{neil2016effective,vaila2019deep} and DNNs such as \cite{neil2016phased,bi2019graph}.

\textbf{\emph{SHD classification.}}
 Spiking Heidelberg Digits (SHD) is a speech dataset in spike format, consists of spoken digits recordings ranging from 0 to 9 in English and German \cite{cramer2019heidelberg}. Spikes are generated by converting audio recordings using artificial inner ear model, resulting 700 spike trains with vary temporal structures. The dataset aims at mimicking the spike activity of biological auditory system, and the spike events exhibit complex spatial temporal distribution. The dataset has 20 classes, and splits into 8156 and 2264 samples for training and testing respectively. A sample of the SHD dataset is shown in Fig. \ref{fig:samples} (b), in which a dot at $(x,y)$ indicates a spike event in $y_{th}$ channel at time $x$. 
 
 We built a spiking MLP to classify this dataset. Network structure is $700 {-} 400{-} 400 {-} 20$. Results are shown in Table \ref{tab:classification}. We achieved $85.69 \%$ accuracy, which is the best in SNN domain. It is noteworthy that \cite{cramer2019heidelberg,yin2020effective} employ explicit recurrent architecture, i.e. neurons within one layer are also connected to each other through synapse connection. Such network topology is difficult to implement on crossbar. Our results show that it is possible to achieve similar performance with simple feedforward structure.

\textbf{\emph{Importance of adaptive threshold.}} The 'HR' in Table \ref{tab:classification} refers to 'hard reset'. We keep the network structure and weights, but swap the neuron to ODE model defined by \eqref{eq:neuron_model}. On N-MNIST dataset, the accuracy dropped to $95.31 \%$, and on SHD dataset, the accuracy dropped to $26.36 \%$. This clearly shows the drawback of widely used hard reset model, as claimed in Section \ref{sec:model}, hard reset clears historical information, therefore it impairs the capability of SNN to process temporal information. The accuracy degradation on SHD is more significant than N-MNIST. We presume that it is because the richness of temporal information in the two datasets are different. \cite{cramer2019heidelberg} claimed that spike timing is essential in SHD dataset. \cite{iyer2018neuromorphic} reported that rate-based model can achieve high accuracy on N-MNIST with data pre-processing, the temporal information is not dominant. Above results justify the capability of our neuron model and the necessity of proposed circuit design, to apply neuromorphic hardware to temporal pattern classification, neural dynamic should be taken into consideration.

\subsection{Spatial Temporal Pattern Association} \label{sec:association}

\begin{figure}[t]
  \centering
  \includegraphics[width=\linewidth]{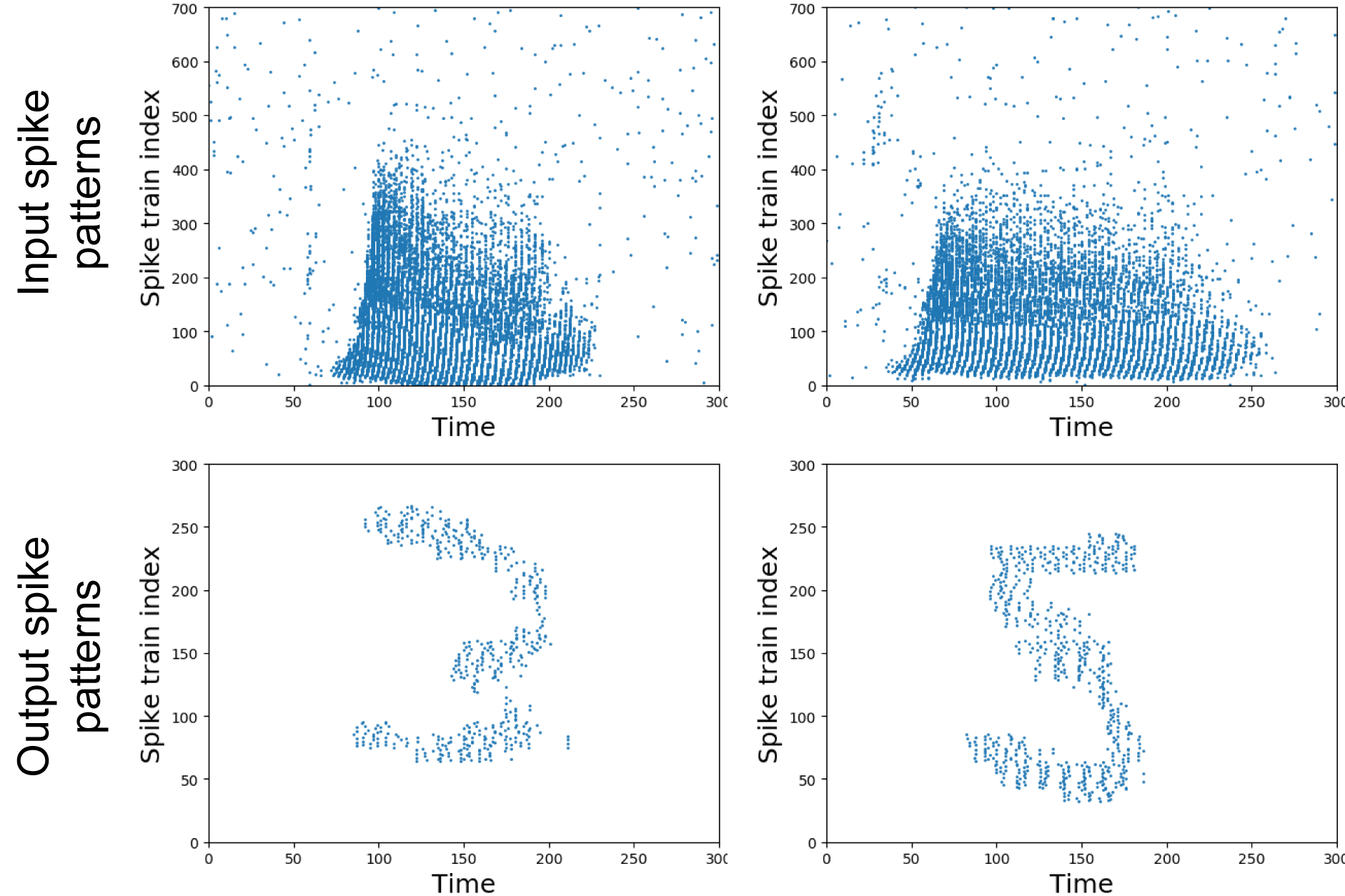}
  \caption{Pattern association input and output samples.}
  \label{fig:association}
\end{figure}

In this task, we build a fully connected network with size 700(input)-500-500-300. The objective is to train the network to produce a specific spatial temporal pattern when given another spatial temporal pattern as input. We randomly pick 1000 samples from SHD dataset as input patterns. Then we convert handwritten digit images representing 0 to 9 to spike patterns by regarding pixel $(x,y)$ as a spike event in $y_{th}$ spike train at time $x$. The input has 700 spike trains of length 300, the target output has 300 spike trains of same length. The network is trained to output corresponding handwritten digits pattern when given a SHD sample. For example, when given the audio representation of digit 3, the network generates spike pattern that resembles handwritten digit 3. Example inputs and outputs are shown in Fig. \ref{fig:association}. The x axis represents time, and y axis represent spike train index, a dot at coordinate $(x,y)$ represents an input/output spike in $y_{th}$ spike train at time $x$. This task demonstrates the advantage of our training algorithm. It can train SNN to learn precise spike timings, enabling novel applications beyond simple static image classification.

\subsection{Hardware}

\begin{figure}[t]
  \centering
  \includegraphics[width=\linewidth]{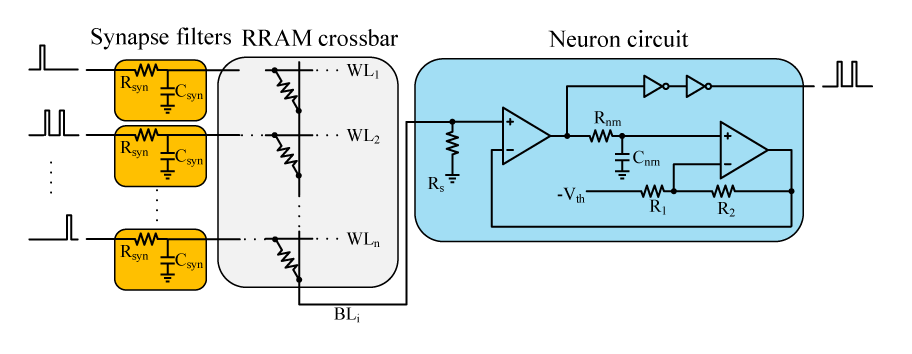}
  \caption{Diagram of synapse filter, RRAM crossbar, and neuron circuit.}
  \label{fig:circuit}
\end{figure}

A circuit implementation was designed and simulated in Cadence Virtuoso using TSMC 1V-65 nm technology node. A diagram of the designed neuron circuit can be seen in Fig.~\ref{fig:circuit}. The synaptic circuit is an RC filter placed at the input of the RRAM array and filters incoming spikes for each word-line. A resistor is placed between the output of bit-line and ground to convert the synaptic current into voltage (PSP). Beyond these devices, the circuit consists of RC filter in the neuron, an operational amplifier that behaves as a comparator, another operational amplifier acting as our bias voltage source ($V_{th}$), and two simple inverters. We chose a physical step size, which corresponds to the width of an input spike, of 10ns. To achieve a similar time constant $\tau$ to that in the simulations, we chose an RC filter with \(R = 4.56k\Omega\) and \(C = 10.14pF\), which gives us the desired 40ns time constant. A bias voltage of 550mV was chosen for an initial experiment to ensure that the neuron would not spike with every input spike.

\begin{figure}[ht!]
  \centering
  \includegraphics[width=\linewidth]{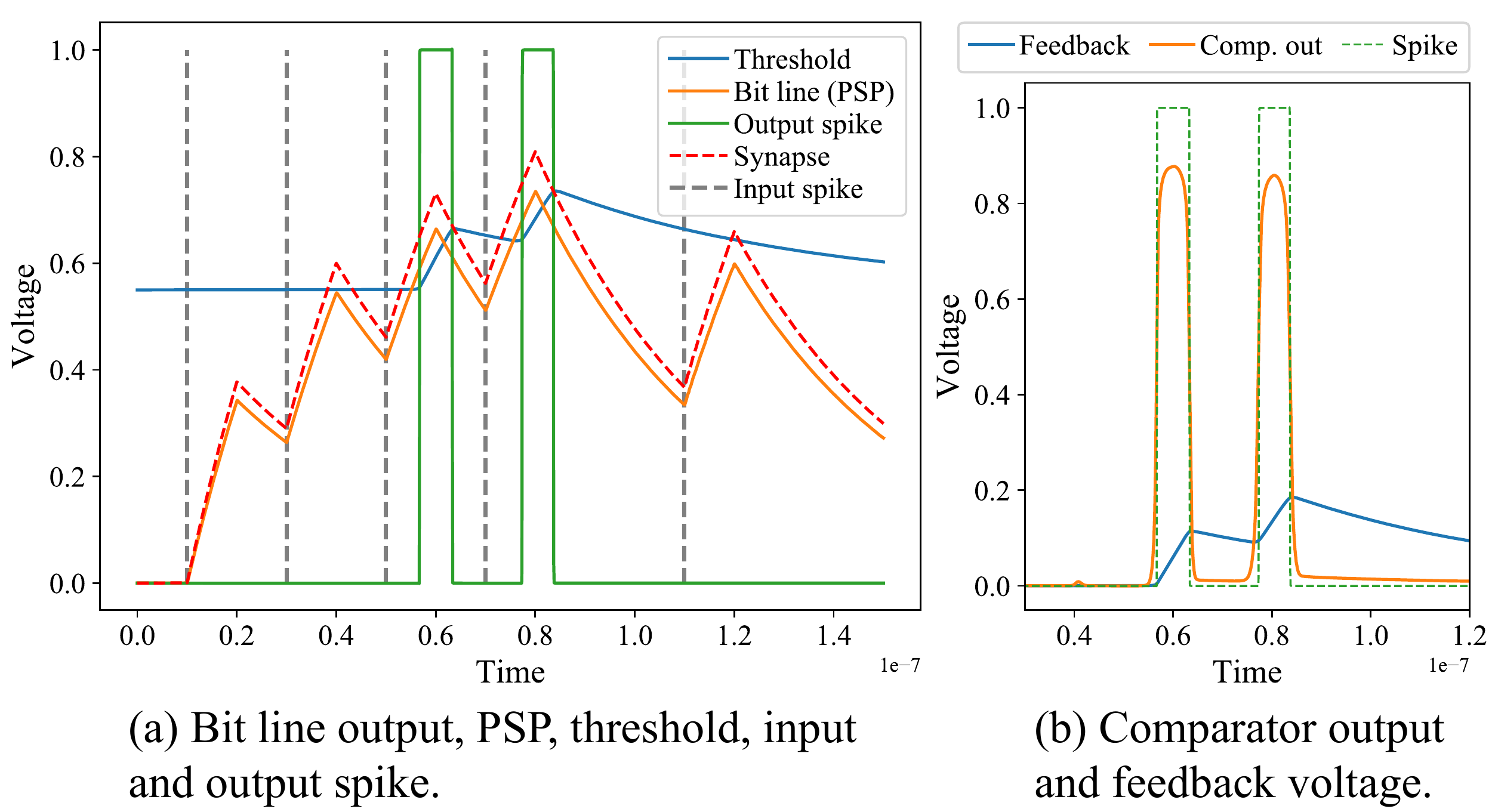}
  \caption{Circuit simulation results.}
  \label{fig:simulation}
\end{figure}

\begin{figure}[t]
  \centering
  \includegraphics[width=0.80\linewidth]{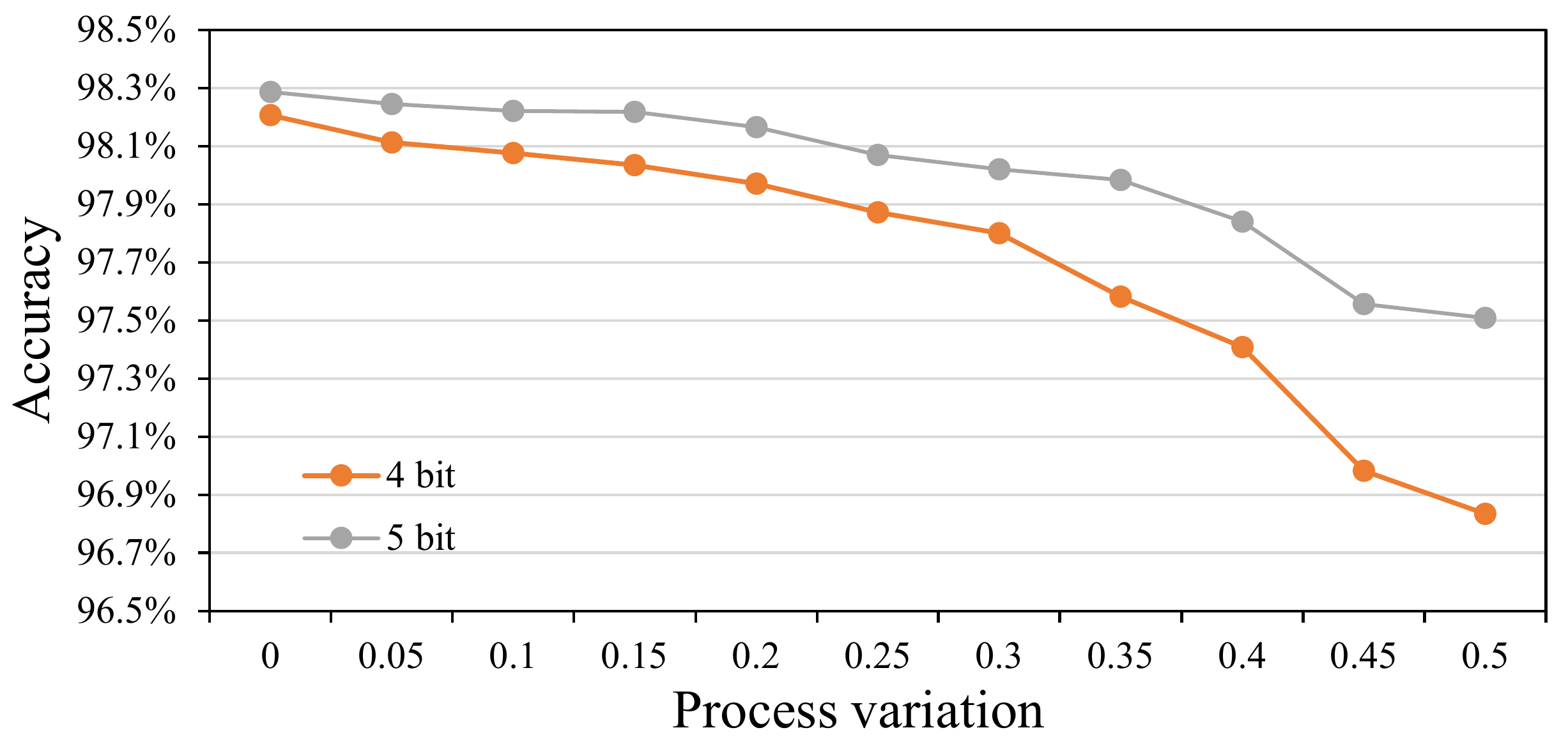}
  \caption{Accuracy under different quantization level and process variation.}
  \label{fig:variation}
\end{figure}

The dynamics of the implemented circuit can be seen in Fig.~\ref{fig:simulation}. A given input spike train is filtered to produce $k(t)$, and the output of the comparator rises as the bit-line output $g(t)$ surpasses the overall threshold $V_{th} + r(t)$. The output of the comparator is filtered to produce $h(t)$, shown as "feedback" in Fig.~\ref{fig:simulation} (b). This feedback voltage goes back to comparator, turning off the comparator as the threshold increases. The threshold decays slowly back to the bias as $h(t)$ decays, but is high enough to prevent a subsequent input spike from inducing an output spike. Thus, the circuit replicates the dynamics demonstrated by the model. Because the output of the first amplifier is non-ideal, shown by yellow curve in Fig.~\ref{fig:simulation}, the inverters are used to generate spike with ideal shape shown by dashed green curve in Fig. \ref{fig:simulation} (b).

We next estimated area and power consumption of the circuit. An arbitrary input sample of 300 time steps with 14 input spikes was provided as input to the circuit, and the threshold bias was adjusted to replicate the neuron activation statistics of our software model. We found that a single neuron and synapse circuit consumed a minimum of 1.067$mW$, a maximum of 1.965$mW$, and an average of 1.11$mW$. Integrating the power over the entire sample duration, the circuit consumed 3.329$nJ$ of total energy. We then calculated the footprint of all devices in the circuit to estimate a total area of about 0.0125$mm^2$ for a single neuron and synapse circuit. Note that these estimates are independent of RRAM array size and energy consumption, and are an estimate of the neurosynaptic dynamics circuitry.

We also simulated the influence of process variation and quantization using the same N-MNIST classification model as in Section \ref{sec:classification}. Results of 4-bit and 5-bit quantization with process variation (resistance deviation) ranging from 0 to 0.5 are shown in Fig. \ref{fig:variation}. Our model shows robustness to quantization and process variation. With 4-bit level precision, 0.2 deviation, we achieved 97.97\% accuracy.

\section{Conclusion}

In this work, we proposed algorithm hardware co-design for memristor based neuromorphic computing systems. We developed a hardware friendly spiking neuron model and corresponding training algorithm to learn complex spatial temporal patterns. The algorithm has achieved competitive performance on commonly used spiking datasets and enables novel applications. The hardware was designed based on proposed model, it exhibits critical neural dynamics including synapse filter effect and adaptive threshold.

\bibliographystyle{abbrv}
\bibliography{conference_101719}

\end{document}